\documentclass[lettersize,journal]{IEEEtran}

\usepackage{cite}
\usepackage{graphicx}
\usepackage{xcolor} 
\usepackage{amsmath} 
\usepackage{booktabs} 
\usepackage{adjustbox}
\usepackage{multirow}

\graphicspath{{./images/}}

\begin{document}
\title{Unified Framework for Identity and Imagined Action Recognition from EEG patterns}

\author{Marco~Buzzelli, Simone~Bianco, and Paolo~Napoletano
\thanks{M. Buzzelli, S. Bianco and P. Napoletano  are with the Department of Informatics Systems and Communication, University of Milano - Bicocca, Italy. e-mail: \{name.surname\}@unimib.it. This work has been submitted to the IEEE for possible publication. Copyright may be transferred without notice, after which this version may no longer be accessible}
}

\markboth{Journal of \LaTeX\ Class Files,~Vol.~14, No.~8, August~2021}%
{Buzzelli \MakeLowercase{\textit{et al.}}: Unified Framework for Identity and Imagined Action Recognition from EEG patterns}

\maketitle

\begin{abstract}
We present a unified deep learning framework for the recognition of user identity and the recognition of imagined actions, based on electroencephalography (EEG) signals, for application as a brain-computer interface.
Our solution exploits a novel shifted subsampling preprocessing step as a form of data augmentation, and a matrix representation to encode the inherent local spatial relationships of multi-electrode EEG signals.
The resulting image-like data is then fed to a convolutional neural network to process the local spatial dependencies, and eventually analyzed through a bidirectional long-short term memory module to focus on temporal relationships.
Our solution is compared against several methods in the state of the art, showing comparable or superior performance on different tasks.
Specifically, we achieve accuracy levels above 90\% both for action and user classification tasks. In terms of user identification, we reach 0.39\% equal error rate in the case of known users and gestures, and 6.16\% in the more challenging case of unknown users and gestures.
Preliminary experiments are also conducted in order to direct future works towards everyday applications relying on a reduced set of EEG electrodes.

\end{abstract}


\IEEEpeerreviewmaketitle

\section{Introduction}

\IEEEPARstart{B}{rain}-computer interface (BCI)
is a term that refers to a wide variety of techniques and technologies, designed to establish a direct communication link between the brain (or, more generally, functional components of the central nervous system) and an external computational device.
BCI, sometimes also referred to as brain-machine interface (BMI), can be decisive in improving the life quality of people affected by motor-impairing disorders, but it can also be employed for entertainment purposes.

A recent review of state of the art methods for BCI-controlled vehicles~\cite{hekmatmanesh2021review} identified electroencephalogram (EEG), electrooculogram (EOG) and electromyogram (EMG) as bio-signals that are commonly used for vehicle control, often relying on deep learning techniques for the identification of imaginary actions.
Among the various technologies that enable BCI, here we focus on the usage of EEG signals as collected by a non-invasive mesh of electrodes to be worn on the subject's head.
BCIs can be applied as a tool to transfer an imagined action (as opposed to a physical action~\cite{ferrari2020personalization,buzzelli2020vision}) to a physical actuator, such as assistive robotic arms~\cite{schiatti2017soft} and spherical exploration robots~\cite{halme1996motion,tomik2012design}.
BCI has also found application in authentication systems, where the user's brain signature can be exploited as a behavioral biometric key for identification and verification.
This has the advantage of allowing authentication for motor-impaired people, or improving the robustness as ``inherent'' component in multi-factor authentication systems.

\begin{figure}[!t]
\centering
\includegraphics[width=\columnwidth]{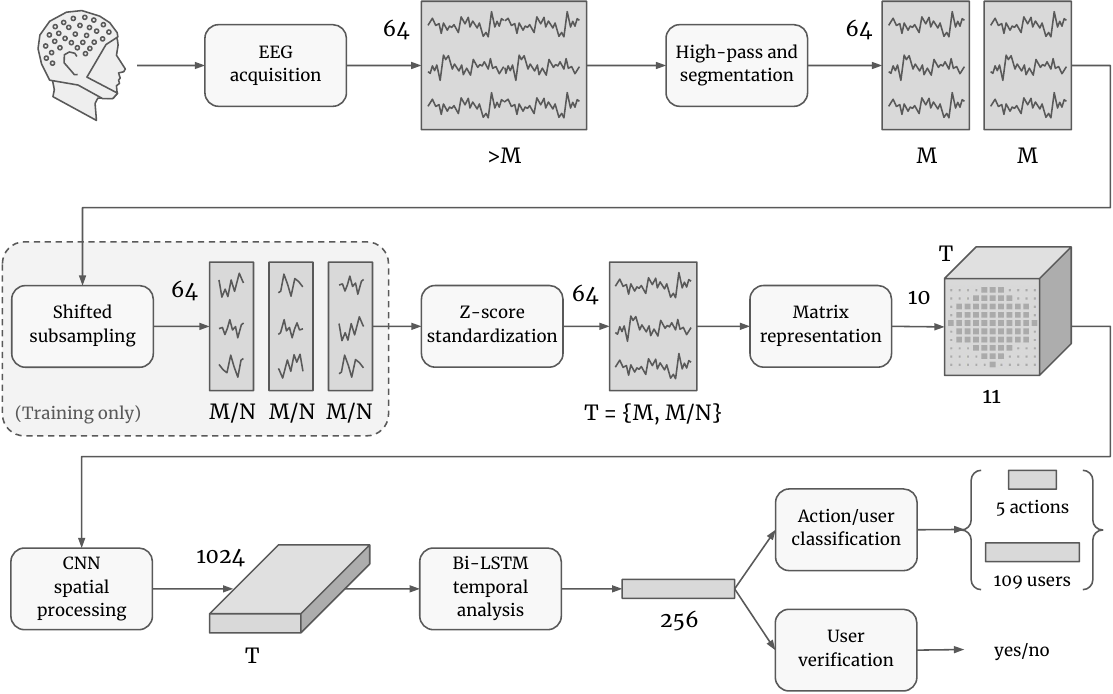}
\caption{Presented pipeline of our unified framework for recognition of user identity and imagined action.}
\label{fig:pipeline}
\end{figure}

In this paper we present a novel approach, based on artificial neural networks, that simultaneously addresses multiple tasks: action recognition, user identification, and user verification, all within a unified framework.
To the best of our knowledge, this is the first use case of the joint exploitation of EEG data for user authentication and action recognition.
Conceptually, we can envision a scenario where the same application technology, such as the aforementioned robotic arms or exploration robots, can be both accessed and controlled with a unique BCI.

Our solution, illustrated in Figure~\ref{fig:pipeline}, is based on the following contributions:
\begin{itemize}
    \item Shifted subsampling preprocessing for data augmentation. This novel procedure is formally defined, and its benefits are experimentally quantified.
    \item Matrix representation for the exploitation of local spatial relationships. Its usage in conjunction with shifted subsampling and our specific architecture leads to results that outperform other solutions also inspired by matrix representation.
    \item Two-stage neural architecture composed of a convolutional backbone for local spatial processing, followed by a bidirectional long-short term memory (Bi-LSTM) head for exploitation of temporal relationships.
    \item Analysis of the extracted neural features, commonly treated as a black box, revealing that features learned for user classification retain some discerning capability that also applies to action recognition.
    \item Sensitivity study on the number and location of electrodes, providing guidelines for future integration in wearable devices.
\end{itemize}
The developed solution is proven to be highly effective, with accuracy levels above 90\% for all classification tasks, and highly efficient, requiring only 0.004 of GPU processing time per each second of EEG data.

The paper is structured as follows.
Section~\ref{sec:related} presents significant works from the scientific literature in terms of EEG representation and interpretation, as well as the existing datasets of EEG data.
Section~\ref{sec:method} describes in detail our proposed solution for action recognition and user identification/verification.
Section~\ref{sec:experiments} defines the adopted experimental protocol and provides in-depth comments and analyses on the experimental results.

\section{Related works}
\label{sec:related}

\subsection{Methods for EEG representation and interpretation}

\subsubsection{Application-agnostic representations}
The scientific literature offers several works related to the analysis of EEG signals, starting from those specifically focused on developing features and representations that can be used for a variety of downstream applications.

Bashivan et al.~\cite{bashivan2016learning}
specifically showed the importance of crafting features that are invariant to inter- and intra-subject differences, as well as to inherent noise associated with the collection of EEG data. They transformed the signals into a sequence of topology-preserving multi-spectral images. They then trained a deep recurrent neural network (RNN), and focused on a task of cognitive load classification for representation learning and assessment.
Inspired by the described data representation, here we introduce a matrix transformation that enables a 2D convolutional processing, while reducing the potential loss of information introduced by the interpolation used in~\cite{bashivan2016learning}.

Kasabov et al.~\cite{kasabov2014neucube}
argued that spiking neural networks (SNN) are particularly suited to the representation of EEG and other types of spatio/spectro temporal brain data, as they are based on the same computational principle, i.e. spiking formation processing.
Based on this observation, they defined a neural architecture (NeuCube) based on a three-dimensional evolving SNN, aiming at approximating the structural and functional
areas of interest of the brain that are related to modeling temporal brain data.
The proposed representation was presented as potentially useful for the discovery of functional pathways from data, as well as the prediction of future brain activities.

Monsy and Vinod~\cite{monsy2020eeg}
devised a novel feature called frequency-weighted power (FWP), which represents the power of a specific frequency band multiplied and integrated by its corresponding power density value. The authors evaluated this representation in the context of user identification, and hypothesized that the improvement with respect to traditional power spectral density (PSD) is due to FWP's dependency on a specific frequency point, as such carrying more discriminative information.

\subsubsection{Application-specific representations}
Parallel to the research within application-agnostic representations which are not designed for any individual use case, a branch of scientific literature focused instead on the development of application-specific representations, such as those aimed at biometric identification.

Fraschini et al.~\cite{fraschini2014eeg}
presented a phase-synchronization approach aimed at recognizing and exploiting personal distinctive brain network organization.
The authors band-pass-filtered the subjects EEG signals, and estimated functional connectivity using the phase lag index. They then reconstructed a weighted network based on the resulting connectivity matrix, and subsequently computed the nodal eigenvector centrality, which was used to assign a level of importance to different neural regions.
According to their findings, resting-state functional brain network topology appears to provide better discriminative power than using only a measure of functional connectivity.

Thomas and Vinos~\cite{thomas2018eeg}
proposed a simple yet effective authentication technique based on cross-correlation values of PSD features of gamma band (30–50 Hz), showing its superior performance when compared to
delta, theta, alpha and beta bands of the signals.

La Rocca et al.~\cite{la2014human}
argued that traditional features, such as power spectrum estimation, which characterize the activity of single brain regions, ignore possible temporal dependencies between the generated signals.
Based on the observation that richer features reside in the functional coupling of different brain regions, the authors proposed the fusion of spectral coherence-based connectivity between such regions.

Subsequently,
Yang et al.~\cite{yang2018task}
identified a potential problem in using the EEG signals captured
during the so-called ``resting phase'', due to the ambiguity of interpretation by the users, leading to uncontrollable heterogeneity in the data.
Additionally, the authors proposed a novel feature based on wavelet transformation to extract identity-specific information.
They adopted an experimental protocol aimed at establishing the difference in performance when training and testing using data collected from different tasks.
In general, it has been proven that proper data representation and encoding are crucial for successful interpretation. To this extent, we will describe our data preprocessing and representation pipeline in Section~\ref{sec:preproc}, which includes encoding the spatial relationship of the sensor mesh into a three-dimensional matrix.

\subsubsection{Neural-network-based solutions}
In more recent years, many researchers migrated to the design and development of application-specific uses of EEG data, leaving the task of feature definition to the training process of neural networks, showing remarkable results.

With respect to action recognition, also referred to as motion imagery,
Chen et al.~\cite{chen2018eeg}
proposed a method that applies multi-task RNNs to learn distinctive features from EEG signals, and exploits the temporal correlation between different frequency channels to improve the recognition of imagined actions.

Dose et al.~\cite{dose2018end}
devised a convolutional neural network (CNN) for imagined action recognition, and extended it for subject-specific adaptation. They conducted an analysis of the learned filters, to provide a level of interpretability on the predicted class.

Zhang et al.~\cite{zhang2019classification}
addressed the classification of left/right hand movement by extracting preliminary features from the time and frequency domain, and subsequently processing them through a LSTM network that exploits an attention mechanism.

Zhang et al.~\cite{zhang2019making}
developed two levels of neural processing: a convolutional and a recurrent neural network. These were used jointly in either a cascade or parallel setup, to explore the space of feature representations.

In terms of user identification and verification,
Sun et al.~\cite{sun2019eeg} also
combined a traditional CNN with an LSTM module to extract data-defined spatio-temporal features of the EEG signals, and showed the empirical improvement in user identification accuracy. The authors also experimented with reducing the number of electrodes to potentially lower the cost of a practical implementation of a user recognition system.

Recently, Lu et al.~\cite{lu2020deep} produced a comprehensive comparison study, which confirmed the effectiveness of preceding an RNN-based temporal analysis by a CNN-based preprocessing of the input signals.
We leverage this finding by designing an original neural architecture structured according to the aforementioned paradigm: jointly exploiting a convolutional backbone for data-driven preprocessing, and a Bi-LSTM head for handling temporal relationships. When combined with our shifted subsampling data augmentation, we are able to fully exploit the task separation and achieve excellent recognition performance for both action recognition and user identification.

\subsection{Datasets of EEG data}
\label{sec:datasets}

The domain of EEG signal processing is associated with a number of datasets released through the years.

In 1990, Keirn and Aunon~\cite{keirn1990new} published the eponymous dataset. For data collection, the electrodes were connected through band-pass analog amplifier filters set at 0.1-100 Hz, and sampled at 250 Hz. All the data was converted with an IBM-AT controlling a Lab Master analog-to-digital converter at 12 bits of accuracy.

In 1999, Beigleiter~\cite{beigleiter1999eeg} released the UCI KDD (University of California Irvine, Knowledge Discovery in Databases) dataset.
The data was collected within the framework of a larger study on the correlation between EEG signals and the genetic predisposition to alcoholism. It contains measurements from 64 electrodes placed on the scalp of the subjects, sampled at 256 Hz for 1 second.

Between 2001 and 2008, four versions of the BCI Competition have been organized~\cite{sajda2003data,blankertz2004bci,blankertz2006bci,tangermann2012review}.
With the final iteration of the challenge, several tasks have been included, such as classifying signals affected by eye movement artifacts, or the direction of wrist movements based on magneto-encephalography, and the associated dataset are characterized by a wide variety of bands and sampling rates.

In 2005 Hunter et al.~\cite{hunter2005australian} published the Australian EEG Database: a web-based dataset of 18,500 EEG records from a regional public hospital throughout a time span of 11 years. A notable feature of this dataset is the variety of users, ranging in age from a premature infant born at 24 weeks gestation, up to 90 year old people.

In 2012 Koelstra et al.~\cite{koelstra2011deap} released DEAP: a database for emotion analysis using physiological signals.
Both the EEG and peripheral physiological signals of 32 participants were recorded while watching 40 one-minute-long excerpts of music videos.
In addition to the recorded signals, user ratings and frontal video recordings were also collected.

\subsubsection{PhysioNet}
The data used in this paper comes from the PhysioNet dataset for motor movement/imagery (MMI)~\cite{www_physionet,goldberger2000physiobank}, collected and processed using the BCI2000 software system~\cite{schalk2004bci2000}.
The EEG data are recorded using 64 electrodes, with a sampling rate of 160Hz, and positioned according to the international 10-10 system (excluding electrodes Nz, F9, F10, FT9, FT10, A1, A2, TP9, TP10, P9, and P10).
The dataset was collected by having 109 human subjects perform a number of guided operations, which involve both the real and imagined movement of body parts in response to visual stimuli.
For the purpose of this work, we focus on the analysis of EEG data collected during the performance of imagined actions only.
The EEG activity recorded during these sessions is then assembled in different ways, to define the various recognition tasks addressed in this paper: action recognition, user identification, and user verification.

The data collection sessions were structured as follows.
Each of the 109 subjects performed 14 so-called ``runs'':
a one-minute baseline run with open eyes, a one-minute baseline run with closed eyes, and three run cycles of the following four tasks:
\begin{enumerate}
    \item Task 1: a visual target is shown on either the left or right side of a screen.
    The subject opens and closes either the left or the right hand (depending on the target position).
    \item Task 2: a visual target is shown on either the left or right side of a screen.
    The subject \textit{imagines} opening and closing either the left or the right hand.
    \item Task 3: a visual target is shown on either the top or bottom side of a screen.
    If the target is on the top, the subject opens and closes both hands.
    If the target is on the bottom, the subject opens and closes both feet (by contraction of the toes).
    \item Task 4: a visual target is shown on either the top or bottom side of a screen.
    If the target is on the top, the subject \textit{imagines} opening and closing both hands.
    If the target is on the bottom, the subject imagines opening and closing both feet.
\end{enumerate}
Each of these four tasks, for each of the three cycles, lasts for 2 minutes, alternating 4 seconds of action to 4.2 seconds of rest.
For all of our experiments, we focused on the usage of EEG data recorded while the subjects were performing imagined actions, or resting.

\section{Proposed method for action and identity recognition}
\label{sec:method}

Based on EEG data relative to motor imagery, we propose a unique solution capable of performing three tasks: action recognition, user identification, and user verification.
The first two tasks are pure classification problems, while the third task can be described as a metric learning problem and it is derived from the second classification model (user identification).
More specifically, user identification refers to a closed set classification scenario, where the authentication module directly outputs a class (user) by processing the input data. As a consequence, this type of authentication can be used in a static environment where the list of possible users is defined once before deployment.
In many practical applications, however, the list of possible users is constantly changing. In these situations, it is common to reformulate the problem in terms of user identification: the authentication module does not directly output a class (user), but compares features from the authenticating user with features from a dynamic list of registered users, eventually returning the most similar match above a rejection threshold. Alternatively, user verification can also be used to compare the authenticating user with a single registered user, as indicated by the ``yes/no'' output in Figure~\ref{fig:pipeline}.
Additional details on the experimental protocol are provided in Section~\ref{sec:problem}.

\subsection{Data preprocessing and representation}
\label{sec:preproc}

The dataset is first preprocessed with a 1Hz high-pass filter, in order to reduce noise, the negative effect of signal artifacts, and power-line interference.

The filtered signals are segmented with a non-overlapping sliding window that spans 80 samples.
For example, 1 second of data, corresponding to 160 samples, is split into two 0.5s windows, composed of 80 samples each.

\begin{figure}[!t]
\centering
\includegraphics[width=.85\columnwidth]{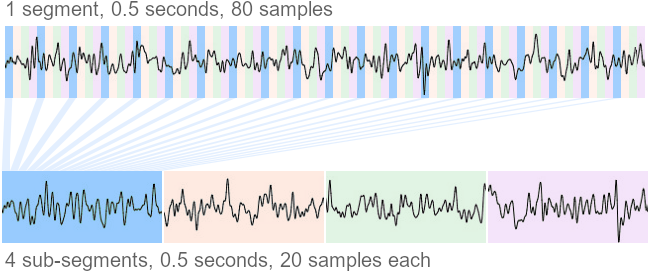}
\caption{Visualization of the subsampling-based data augmentation technique. The same 0.5 second-long segment (80 samples) is replaced with 4 sub-segments obtained by varying the sampling temporal shift.}
\label{fig:sampling}
\end{figure}
A novel shifted subsampling preprocessing is then applied on segmented data as a form of data augmentation, by sampling each segment multiple times at a fixed interval with varying temporal shift. Let:
\begin{equation}
    S = \{ s_1, s_2, ..., s_M\}
\end{equation}
be a segment $S$ composed of $M$ samples.
The corresponding data augmentation produces $N$ sub-segments $S_{(n)}$:
\begin{equation}
    S_{(n)} = \{ s_i : i = n+jN \}
\end{equation}
with $j=\{0, 1, ..., \lfloor M/N \rfloor \}$ and $n = \{1, 2, ..., N\}$.
Preliminary experiments showed that the joint use of segmentation and subsampling-based data augmentation with $N=4$ allows for improved performance in all tasks.
As an example, the described procedure leads to 20-samples-long data sub-segments extracted from 0.5 seconds of raw signal, as visualized in Figure~\ref{fig:sampling}.

Z-score standardization is then applied to the information from each electrode considered individually, by rescaling the data distribution so that the mean of observed values is 0 and the standard deviation is 1.

\begin{figure}[!t]
\centering
\includegraphics[width=\columnwidth]{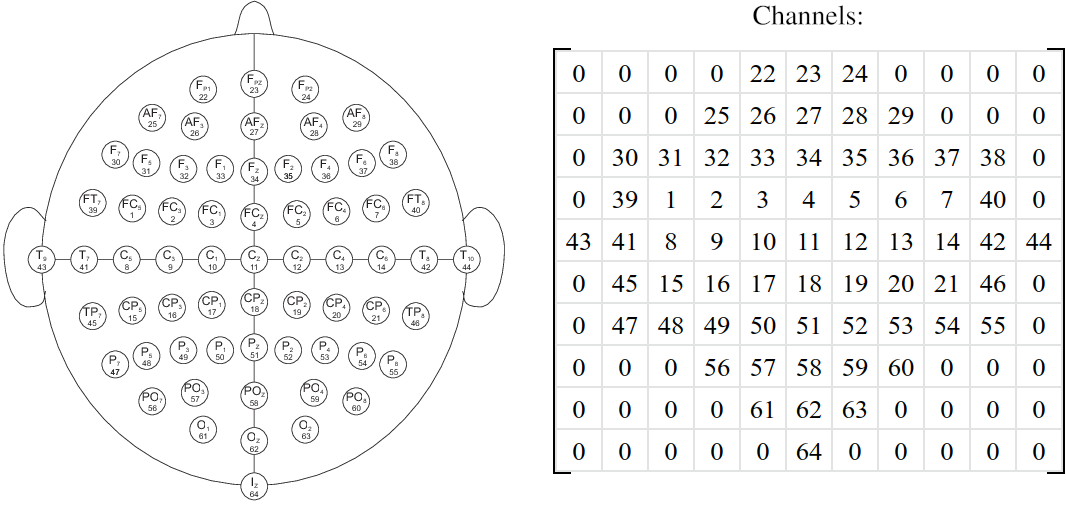}
\caption{Channel correspondence of our matrix representation mapping, where each electrode's value at time $t$ is encoded as a pixel in a $10\times 11$ matrix.}
\label{fig:mesh}
\end{figure}

Finally, the relative position of the electrodes is encoded in a matrix representation that preserves the three-dimensional nature of the underlying data, composed of 10 rows, 11 columns, and 20 channels.
The rows and columns directly encode the spatial distribution of the electrodes on the user's scalp, following the mapping visualized in Figure~\ref{fig:mesh}, whereas the channels encode the temporal information after the segmentation and subsampling procedure previously described.
It can be noted that our matrix representation mapping involves a discrete relocation of the exact input values~\cite{zhang2019making,lu2020deep}. Differently, the approach originally introduced by Bashivan et al.~\cite{bashivan2016learning} consists of a continuous projection from a 3D surface to a 2D image, relying on values interpolation and thus possibly causing loss of information.

\subsection{Neural architecture}
\label{sec:ourmodel}

We define a hybrid neural model composed of a CNN that processes the input 3D data with a pyramid representation, followed by a RNN to specifically exploit the temporal component of the input signal, using a Bi-LSTM module. Alternatively, this sequential structure could be also replaced with a parallel one~\cite{zhang2019making}, although such investigation is left for future developments.
The overall architecture is illustrated in Table~\ref{tab:architecture}.

\begin{table}[t]
\caption{Architecture of the proposed multi-purpose neural network. The dimensionality and output of the fully-connected layer depend on the task at hand (5 for action recognition, and 109 for user identification).}
\label{tab:architecture}
\begin{tabular}{lccccc}
\toprule
Layer type & Layer characteristics & \multicolumn{4}{c}{Output shape} \\
\midrule
Input &  & 20 & 10 & 11 & 1 \\
Time distr. convolution & $3 \times 3 \times 1 \times 32$ & 20 & 10 & 11 & 32 \\
Batch normalization &  & 20 & 10 & 11 & 32 \\
ELU activation &  & 20 & 10 & 11 & 32 \\
Time distr. convolution & $3 \times 3 \times 32 \times 64$ & 20 & 10 & 11 & 64 \\
Batch normalization &  & 20 & 10 & 11 & 64 \\
ELU activation &  & 20 & 10 & 11 & 64 \\
Time distr. convolution & $3 \times 3 \times 64 \times 128$ & 20 & 10 & 11 & 128 \\
Batch normalization &  & 20 & 10 & 11 & 128 \\
ELU activation &  & 20 & 10 & 11 & 128 \\
Time distr. convolution & $3 \times 3 \times 128 \times 256$ & 20 & 10 & 11 & 256 \\
Batch normalization &  & 20 & 10 & 11 & 256 \\
ELU activation &  & 20 & 10 & 11 & 256 \\
Time distr. flatten &  & 20 & \multicolumn{3}{c}{28160} \\
Fully-connected & $28160 \times 1024$ & 20 & \multicolumn{3}{c}{1024} \\
Dropout &  & 20 & \multicolumn{3}{c}{1024} \\
Batch normalization &  & 20 & \multicolumn{3}{c}{1024} \\
ELU activation &  & 20 & \multicolumn{3}{c}{1024} \\
Bi-LSTM &  & \multicolumn{4}{c}{256} \\
Fully-connected& $256 \times \{5,109\}$ & \multicolumn{4}{c}{$\{5,109\}$} \\
\bottomrule
\end{tabular}
\end{table}

The data dimensionality is represented as (temporal$\_$index, sensor$\_$row, sensor$\_$column, channel), and the input size $(20, 10, 11, 1)$ is obtained by reordering the 3D matrix representation described in Section~\ref{sec:preproc}.
A series of four convolutional blocks is then defined to sequentially process the data, by enlarging the channel dimension without restricting the spatial dimensions.
Each block is composed of:
\begin{enumerate}
    \item A time-distributed convolutional layer, which applies convolution independently to each element defined in the first dimension.
    \item A batch normalization layer, which stabilizes the learning procedure by standardizing the input activations.
    \item An exponential linear unit (ELU) activation layer, which applies a smooth transition to handle negative inputs, compared to the traditional rectified linear unit.
\end{enumerate}

The resulting activation is flattened, while still preserving the temporal dimension, into a single high-dimensionality vector (28160 elements per temporal index).
This is then processed with a sequence of: fully-connected layer, dropout, batch normalization, ELU activation, thus producing a 1024-long vector for each temporal index.

The temporal component is finally exploited and processed through a Bi-LSTM, whose hidden layer is internally duplicated and used to handle two versions of the input data: in forward and reverse temporal order.
The resulting activation, a 256-long vector, is eventually mapped to the problem size (5 output classes for action recognition, or 109 classes for user identification) by means of one last fully-connected layer.
The learning process is guided by a softmax cross-entropy loss, comparing the output vector with the index class provided as ground truth annotation.
The network is trained for a total of 80 epochs, with a 0.0001 learning rate, and batch size equal to 32.

\section{Experiments}
\label{sec:experiments}

\subsection{Experimental protocol}
\label{sec:problem}

Three macro tasks have been defined on the basis of the PhysioNet EEG data recorded during the actions described in Section~\ref{sec:datasets}:
\begin{enumerate}

\item Action recognition - The five classes of interest are:
\begin{itemize}
    \item Resting state with closed eyes
    \item Imagined motion of the left fist
    \item Imagined motion of the right fist
    \item Imagined motion of both fists
    \item Imagined motion of both feet
\end{itemize}
Experiments related to action recognition are divided into two subcategories:
\begin{enumerate}
    \item Inter-subject action recognition. 103 out of the 109 available users are selected, excluding 6 subjects with corrupted data~\cite{eva2015comparison,szczuko2017real,zhang2019classification}.
    10-fold cross-validation with stratified random sampling is applied to balance the data distribution, selecting at each fold 90\% of the data for training, and 10\% for test. Both random split and split-by-subject experiments are implemented, to provide a broad view of our solution’s performance in different conditions.
    Temporal windows spanning 1 second of signals have been used.
    \item Intra-subject action recognition. 20 users are selected. Each user is evaluated independently, splitting their data as 90\% for training and 10\% for test.
    Temporal windows spanning 0.25 seconds of signals have been used, in order to virtually increase the available data for each user.
\end{enumerate}

\item User identification - The 109 human subjects are interpreted as classes in a user classification task.
The data is randomly split as 90\% for training/validation (internally 75\% and 25\% respectively), and 10\% for test, following the experimental setup from existing works in the state of the art~\cite{sun2019eeg}.
Temporal windows spanning 1 second of signals have been used.

\item User verification - We optimize a threshold to determine whether any two ``bio-signature'' signals belong to the same human subject.
The signals are described in terms of 256-dimensional neural features using the model trained for user identification, and each neural-encoded signal is compared against all other signals using the Euclidean distance.
The resulting distance matrix is used to compute a receiver operating characteristic curve (considering a valid match when the most similar feature corresponds to the same user) and to consequently evaluate the equal error rate (EER) as the threshold selection criterion.\\
We experimentally verify how the system performance depends on the gesture being imagined, as well as the impact of having an unknown subject at test time, for a total of four experimental combinations:
\begin{enumerate}
    \item Gesture-independent, known users (GI-KU):\\
    The user classification model used for feature extraction is trained on data from all 109 users, using all imagined gestures indiscriminately.
    \item Gesture-independent, unknown users (GI-UU):\\
    The model is trained on data from 99 users, using all imagined gestures indiscriminately. The distance-based verification is performed on the remaining 10 users.
    \item Gesture-dependent, known users (GD-KU):\\
    The model is trained on data from all 109 users, considering the five gestures one at a time. The results from the five gestures are eventually averaged.
    \item Gesture-dependent, unknown users (GD-UU):\\
    The model is trained on data from 99 users, considering the five gestures one at a time. The distance-based verification is performed on the remaining 10 users. The results from the five gestures are eventually averaged.
\end{enumerate}
In all cases, temporal windows spanning 1 second of signal have been used.

\end{enumerate}

\subsection{Results and discussion}

The results for the action recognition task are reported in Table~\ref{tab:ar_inter} for inter-subject evaluation and in Table~\ref{tab:ar_intra} for intra-subject evaluation.
The accuracy for intra-subject action recognition is expectedly higher, suggesting the importance of subject-specific features, although the two evaluations are performed by construction on different test sets, with one user at a time in the latter case. Nonetheless, the more challenging task of inter-subject action recognition produces excellent performance, reaching 93.89\% accuracy with random split, and 85.73\% on a split-by-subject scenario.

The performance of several other methods is also reported, although it should be noted that the extreme heterogeneity in experimental conditions, despite all methods being evaluated on the same dataset, does not allow for a direct comparison. To this extent, the number of classes of interest for action recognition, as well as the number of involved users, are also reported in the table for reference.
For example, the best configurations analyzed in the study by Lu et al.~\cite{lu2020deep} return a higher accuracy for inter-subject action recognition, which is however evaluated on a limited subset of 20 users.
The experiment on inter-subject action recognition is also used as a benchmark for quantifying the impact of shifted subsampling as data augmentation operation: by fixing the temporal window for a fair comparison, we compare our solution from Table~\ref{tab:ar_inter} with two alternatives:
\begin{enumerate}
    \item We use the same sampling frequency, but reduce the sample size by further segmentation.
    \item We use a fraction of the sampling frequency.
\end{enumerate}
The two experiments achieve, respectively, 80.84\% and 89.11\% accuracy, compared to the 93.89\% obtained with shifted subsampling. This indicates that the proposed technique successfully exploits the redundancy of a high sampling frequency, by redistributing its information into separate training examples.

\begin{table}[!t]
    \centering
    \caption{Quantitative evaluation for the task of 5-class action recognition, evaluated in the inter-subject scenario, where multiple users are considered simultaneously at training and test time.}
    \label{tab:ar_inter}
    \begin{adjustbox}{width=\columnwidth}
    \begin{tabular}{lccccc}
    \toprule
    Method & Accuracy & Precision & Recall & Actions & Users \\
    \midrule
    Ours (random split) & 93.89\% & 95.90\% & 91.80\% & 5 & 103 \\
    Ours (split by subject) & 85.73\% & - & - & 5 & 103 \\
    Lu et al. 2020~\cite{lu2020deep} \scriptsize{(CNN)} & 96.47\% & - & - & 5 & 20 \\
    Lu et al. 2020~\cite{lu2020deep} \scriptsize{(RNN)} & 92.95\% & - & - & 5 & 20 \\
    Lu et al. 2020~\cite{lu2020deep} \scriptsize{(TCN)} & 97.89\% & - & - & 5 & 20 \\
    Zhang et al. 2019-I~\cite{zhang2019classification} & 83.20\% & 83.70\% & 82.20\% & 3 & 103 \\
    Chen et al. 2019~\cite{chen2019distributionally} & 77.01\% & 73.85\% & 75.77\% & 3 & 108 \\
    Zhang et al. 2019-II~\cite{zhang2019making} & 98.31\% & - & - & 5 & 108 \\
    Donovan et al. 2018~\cite{donovan2018motor} & 79.44\% & 79.66\% & 79.90\% & 2 & 40 \\
    Chen et al. 2018~\cite{chen2018eeg} & 97.86\% & - & - & 5 & 10 \\
    Zhang et al. 2017~\cite{zhang2017multi} & 79.40\% & 79.91\% & 78.10\% & 5 & 20 \\
    Bashivan et al. 2016~\cite{bashivan2016learning} & 67.31\% & - & - & 5 & 108 \\
    Kasabov 2014~\cite{kasabov2014neucube} & 80.64\% & - & - & 5 & 108 \\
    \bottomrule
    \end{tabular}
    \end{adjustbox}
\end{table}

\begin{table}[!t]
    \centering
    \caption{Quantitative evaluation for the task of 5-class action recognition, evaluated in the intra-subject scenario, where the model is trained and tested independently on the data from each user.}
    \label{tab:ar_intra}
    \begin{tabular}{lccc}
    \toprule
    Method & Accuracy & Actions & Users \\
    \midrule
    Ours & 98.43\% & 5 & 20 \\
    Lu et al. 2020~\cite{lu2020deep} (RNN) & 84.48\% & 5 & 20 \\
    Lu et al. 2020~\cite{lu2020deep} (CNN) & 91.28\% & 5 & 20 \\
    \bottomrule
    \end{tabular}
\end{table}

The results for user identification and for user verification are presented in Table~\ref{tab:ui} and Table~\ref{tab:uv} respectively, with high accuracy and low EERs.
As expected, the EER is higher when testing on unknown users (UU as compared to KU), and it is also higher when assessing a gesture-independent setup (GI as compared to GD), corroborating the findings of the experiments in action recognition.
In terms of comparison with existing methods, note that there is a small intersection of methods that were tested for both user verification and user identification, namely the solutions by Yang et al.~\cite{yang2018task}, Sun et al.~\cite{sun2019eeg}, and Monsy et al.~\cite{monsy2020eeg}.
The considerations about the heterogeneity of experimental conditions, however, also apply in this context, where different methods applied a different selection and organization of the data coming from the PhysioNet dataset.
In terms of user verification, specifically, we classified the existing methods based on the combination described in Section~\ref{sec:problem}, related to gesture independent/dependent (GI/GD) assessment over known/unknown users (KU/UU). Nonetheless, the underlying set of actions highlights how a direct comparison is not possible.

The classification of one second of EEG data, whether in terms of action or user identity recognition, requires 0.004 seconds of GPU time on an NVIDIA Tesla T4, or 0.104 seconds of CPU time on an Intel Xeon CPU @ 2.20GHz, allowing for fast processing. 

\begin{table}[!t]
    \centering
    \caption{Quantitative evaluation for the task of user identification.}
    \label{tab:ui}
    \begin{tabular}{lcc}
    \toprule
    Method & Accuracy & Users \\
    \midrule
    Ours & 99.98\% & 109 \\
    Alyasseri et al. 2020~\cite{alyasseri2020person} & 96.05\% & 109 \\
    Monsy et al. 2020~\cite{monsy2020eeg} & 99.96\% & 109 \\
    Sun et al. 2019~\cite{sun2019eeg} (1D-conv LSTM) & 99.58\% & 109 \\
    Sun et al. 2019~\cite{sun2019eeg} (CNN) & 98.87\% & 109 \\
    Sun et al. 2019~\cite{sun2019eeg} (LSTM) & 96.39\% & 109 \\
    Jin et al. 2019~\cite{jin2019eeg} & 98.55\% & 100 \\
    Yang et al. 2018~\cite{yang2018task} & 99.00\% & 108 \\
    Singh et al. 2015~\cite{singh2015eeg} & 100.00\% & 109 \\
    La Rocca et al. 2014~\cite{la2014human} & 100.00\% & 108 \\
    \bottomrule
    \end{tabular}
\end{table}

\begin{table*}[!t]
    \centering
    \caption{Quantitative evaluation for the task of user verification. The Equal Error Rate (EER) identifies the optimal threshold for intruder detection, and lower is better. Combinations of gesture independent/dependent (GI/GD) assessment over known/unknown users (KU/UU) are considered.}
    \label{tab:uv}

    \begin{tabular}{lccccccc}
    \toprule
    \multicolumn{1}{c}{} & EER & EER & EER & EER &  &  &  \\
    \multicolumn{1}{l}{\multirow{-2}{*}{Method}} & GI-KU & GI-UU & GD-KU & GD-UU & \multirow{-2}{*}{Accuracy} & \multirow{-2}{*}{Actions} & \multirow{-2}{*}{Users} \\
    \midrule
    Ours & 1.07\% & 6.16\% & 0.39\% & 3.88\% & 99.98\% & 5 & 109 (10) \\
    Fraschini et al. 2014~\cite{fraschini2014eeg} & 4.40\% & - & - & - & 92.60\% & 2 & 109 \\
    Yang et al. 2018~\cite{yang2018task} & 4.50\% & - & - & - & 99.00\% & 4 & 108 \\
    Sun et al. 2019~\cite{sun2019eeg} \scriptsize{(1D-LSTM)} & 0.41\% & - & - & - & 99.58\% & 5 & 109 \\
    Sun et al. 2019~\cite{sun2019eeg} \scriptsize{(CNN)} & 1.12\% & - & - & - & 98.87\% & 5 & 109 \\
    Sun et al. 2019~\cite{sun2019eeg} \scriptsize{(LSTM)} & 3.59\% & - & - & - & 96.39\% & 5 & 109 \\
    Monsy et al. 2020~\cite{monsy2020eeg} & - & - & 0.39\% & - & 99.96\% & 2 & 109 \\
    Thomas et al. 2018~\cite{thomas2018eeg} & - & - & 0.80\% & - & - & 2 & 109 \\
    Crobe et al. 2016~\cite{crobe2016minimum} & - & - & 13.05\% & - & - & 2 & 109 \\
    Jijomon et al. 2018~\cite{jijomon2018eeg} & - & - & 1.07\% & - & - & 2 & 109 \\
    Schons et al. 2017~\cite{schons2017convolutional} & - & - & 0.19\% & - & - & 2 & 109 \\
    Mota et al. 2021~\cite{mota2021deep} & - & 0.27\% & - & 0.46\% & - & 2 & 109 \\
    \bottomrule
    \end{tabular}
\end{table*}

\subsection{Features visualization}
\label{sec:tsne}
\begin{figure}[!t]
\centering
    \includegraphics[width=1\columnwidth]{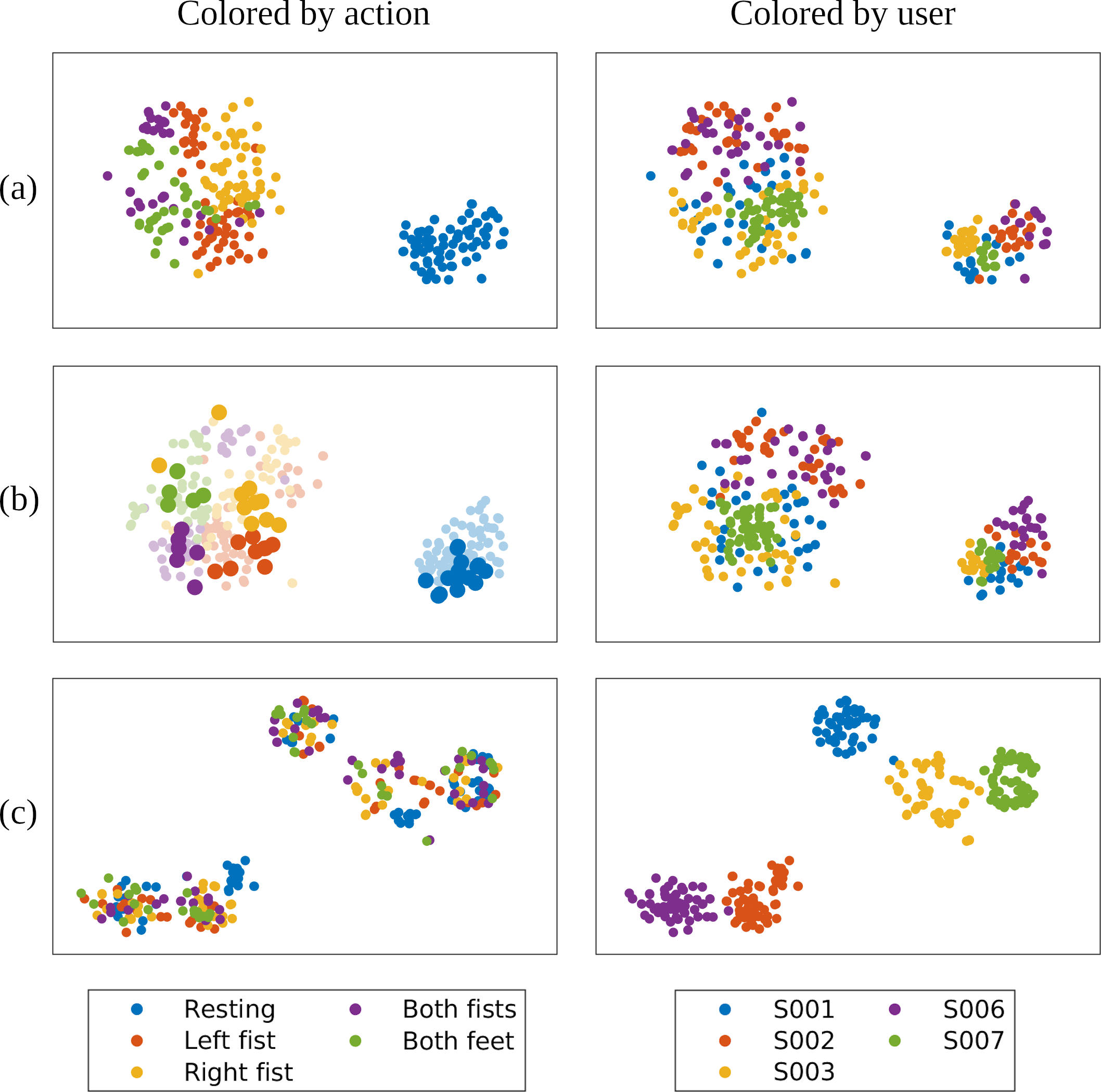}
    \caption{t-SNE projection of neural features obtained after training respectively for: action recognition inter-subject (a), action recognition intra-subject (b), and user identity recognition (c). Data points are colored based on the corresponding action class (left), or user class (right).}
    \label{fig:tsne}
\end{figure}

In this section we present a qualitative comparison of the neural features extracted by the different described classification models: action recognition (inter- and intra- subject) and user identity recognition.
Specifically:
\begin{enumerate}
    \item We train the three classification models.
    Compared to the full experiments reported in the tables, here we restrict to data coming from a subset of five subjects (S001, S002, S003, S006, S007) in order to aid visualization.
    The intra-subject classification is only trained on S001.
    \item We extract the three corresponding feature sets from a common test dataset of readings from the five subjects.
    \item We use t-distributed stochastic neighbor embedding (t-SNE)~\cite{van2008visualizing} to project the extracted features into a two-dimensional embedding space.
    \item We plot the low-dimensional features, coloring each data point by the corresponding action class or user class.
    In the case of intra-subject action recognition, we highlight the test data points that pertain to the same subject used for training.
\end{enumerate}
The results are visualized in Figure~\ref{fig:tsne}.

The data separation performed by both action-based features (rows (a) and (b)) is well correlated with the action-based class subdivision (left column). It is particularly interesting to notice how the ``resting'' class is easily separated from the other actions. When focusing on the specific subject that was used in intra-subject training, however, the overall class separation appears more effective, as highlighted in row (b), left column.
The separation of user classes based on action features is suboptimal (rows (a) and (b), right column), although it is interesting to observe the formation of subject-specific classes, such as the isolation of subject S007.
Conversely, features trained for user identity recognition (row c) provide a close-to-perfect separation of subject classes, in line with the reported classification performance, but they are ineffective for action discrimination.

\subsection{Sensitivity to electrode number and placement}
\label{sec:subset}

\begin{figure}[!t]
\centering
\begin{tabular}{cc}
\includegraphics[width=0.35\columnwidth]{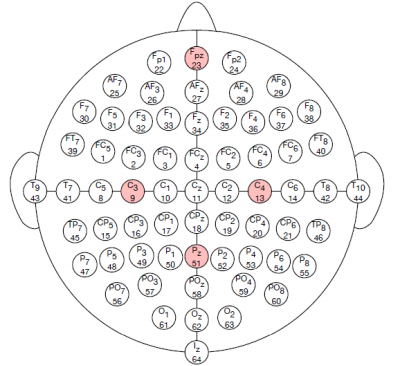}
& 
\includegraphics[width=0.35\columnwidth]{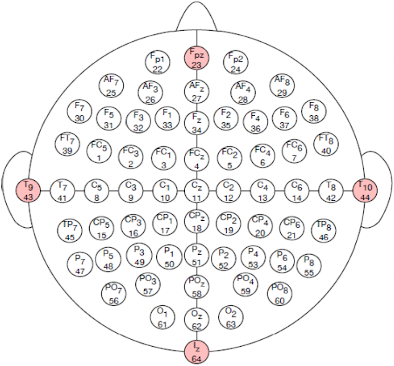}
\\
(a) Enobio 8 (4)
&
(b) EEGlass 4
\\
\includegraphics[width=0.35\columnwidth]{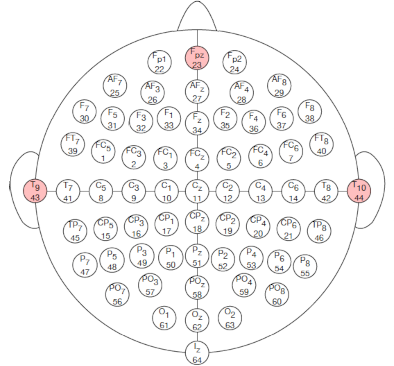}
& 
\includegraphics[width=0.35\columnwidth]{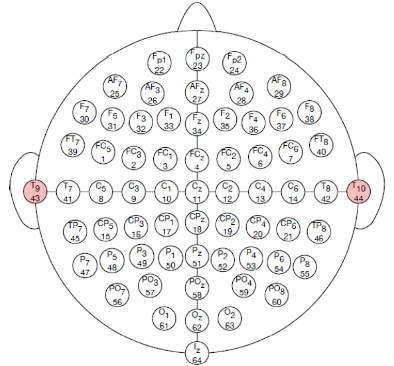}
\\
(c) EEGlass 3
&
(d) EEGlass 2
\end{tabular}
\caption{Electrodes subsets for different configurations.}
\label{fig:subsets_setup}
\end{figure}

In this section we study the sensitivity of our recognition framework to variations in the number and placement of electrodes.
Specifically, this analysis is designed to investigate the possibility of strategically positioning the electrodes in locations that can be accessed by wearable devices, such as smart headbands, smart hats, or smart glasses.
We considered the following configurations, illustrated in Figure~\ref{fig:subsets_setup}:
\begin{itemize}
    \item Enobio 8~\cite{www_enobio8}. 4 electrodes: Fpz, C3, C4 and Pz. 
    \item EEGlass 4~\cite{vourvopoulos2019eeglass}. 4 electrodes: Fpz, T9, T10 and lz. 
    \item EEGlass 3~\cite{vourvopoulos2019eeglass}. 3 electrodes: Fpz, T9, T10. 
    \item EEGlass 2~\cite{vourvopoulos2019eeglass}. 2 electrodes: T9, T10. 
\end{itemize}

\begin{figure}[!t]
\centering
\footnotesize{
\begin{tabular}{cc}
Action Recognition & User Identification \\
\includegraphics[width=0.46\columnwidth]{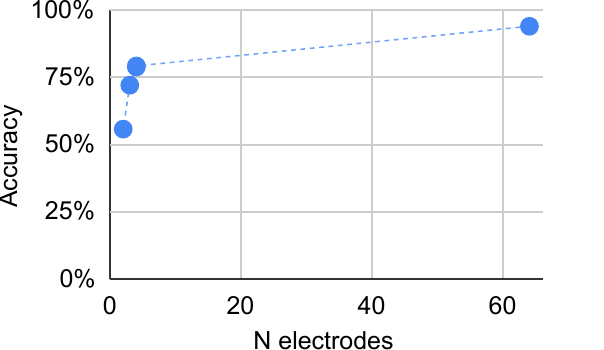}
&
\includegraphics[width=0.46\columnwidth]{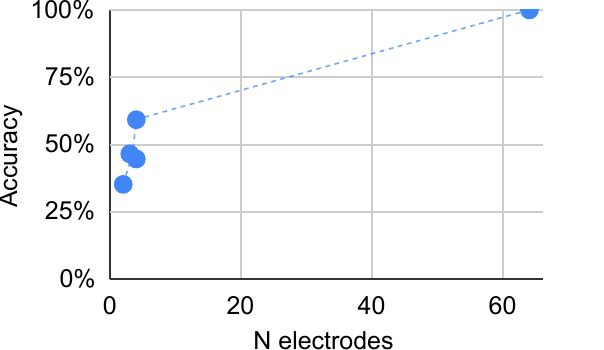} \\
\begin{tabular}{lc}
\toprule
Configuration & Accuracy \\
\midrule
Full (64) & 93.89\% \\
Enobio 8 (4) & 79.14\% \\
EEGlass 4 (4) & 78.76\% \\
EEGlass 3 (3) & 71.99\% \\
EEGlass 2 (2) & 55.66\% \\
\bottomrule
\end{tabular}
& 
\begin{tabular}{lcc}
\toprule
Configuration & Accuracy \\
\midrule
Full (64) & 99.98\% \\
Enobio 8 (4) & 44.55\% \\
EEGlass 4 (4) & 59.15\% \\
EEGlass 3 (3) & 46.47\% \\
EEGlass 2 (2) & 35.13\% \\
\bottomrule
\end{tabular}
\end{tabular}
}
\caption{Action recognition (left) and User identification (right) accuracy with different configurations, by varying the number of electrodes.}
\label{fig:subsets_results}
\end{figure}

All these experiments were re-trained and run with a temporal window spanning 0.5 seconds. 
Figure~\ref{fig:subsets_results} presents a visualization of the decline in performance that results from reducing the number of electrodes, starting from the total of 64 in the ``Full'' configuration.
A good compromise between the number of electrodes and the accuracy of both tasks, appears to be the EEGlass 4 configuration, achievable with a smart headband or hat.
When tested for user verification, this particular configuration yields an EER of 21.89\% for the GD-KU experiment, and 27.63\% for the GI-UU experiment, thus one to two orders of magnitude higher than the full 64-electrode configuration.
This decline in performance suggests that the electrodes-subset configuration is not ideal for user identification, but further investigation is left for future work.

\section{Conclusion}

We have presented a unified neural framework for action recognition, user identification and user verification, based on data collected from EEG signals.
Our solution exploits a shifted subsampling preprocessing to synthetically increase the available training data, and relies on a matrix representation to encode the electrodes pattern into an image-like data structure, which is first processed by a convolutional neural architecture, and finally analyzed by a Bi-LSTM architecture.
The proposed method appears to be effective both in action recognition and user identification, achieving accuracy always well above 90\%.
A comparative study with methods from the state of the art confirms the validity of these results, although a direct comparison is not applicable due to the spread heterogeneity of testing configurations.
User verification has been evaluated in four different setups, in order to provide a full-spectrum baseline for future comparisons, ranging from 0.39\% EER in the gesture-dependent, known user setup, up to 6.16\% EER in the gesture-independent, unknown user setup.

The analysis conducted on features learned for various tasks, highlighted the unexpected action-discriminative power of neural features learned for user identification. This suggests a direction of further research in the domain of transfer learning and semi-supervised learning, the latter successfully applied to multimodal action recognition from wearable sensors~\cite{chen2019semisupervised}, as well as recognition in video sequences~\cite{luo2017adaptive}.

Finally, we have presented a preliminary speculative investigation using a subset of the available electrodes.
The results provide a first broad picture of the decline in performance that stems from manually-selected electrode configuration.
In the future, we will consider genetic algorithms and other meta heuristics~\cite{bakurov2021general} for the automated selection of electrode subsets.


\section*{Acknowledgment}
The authors would like to thank students Andrea Carubelli and Mirko Rima for the support in collecting experimental evidence.

\ifCLASSOPTIONcaptionsoff
  \newpage
\fi

\bibliographystyle{IEEEtran}
\bibliography{biblio}

\end{document}